# Computerized Stokes analysis of optically active polymer films

Georgi Georgiev[a)1,2] and Thomas Slavkovsky[1]

[1]Department of Natural Sciences, Assumption College, Worcester, MA, USA 01609

[2]Physics and Astronomy Department, Tufts University, Medford, MA, USA 02155

**ABSTRACT**

Optics labs are an integral part of the advanced curriculum for physics majors. Students majoring in other disciplines, like chemistry, biology or engineering rarely have the opportunity to learn about the most recent optical techniques and mathematical representation used in today's science and industry optics. Stokes analysis of polarization of light is one of those methods that are increasingly necessary but are seldom taught outside advanced physics or optics classes that are limited to physics majors. On the other hand biology and chemistry majors already use matrix and polarization techniques in the labs for their specialty, which makes the transition to matrix calculations seamless. Since most of the students in those majors postpone their enrollment in physics, most of the registered in those classes are juniors and seniors, enabling them to handle those techniques. We chose to study polymer samples to aid students majoring in other disciplines, especially chemistry and engineering, with understanding of the optical nature of some of the objects of their study. The argument in this paper is that it is advantageous to introduce Stokes analysis for those students and show a lab developed and taught for several years that has successfully, in our experience, done that. Measurements of oriented and unoriented polymer samples are discussed to demonstrate to students the effects of the molecular polarizability on the sample birefringence and the anisotropic Fletcher indicatrix in general.

## I. INTRODUCTION

Stokes analysis provides an integrated approach to learning physics, mathematics, engineering and computer science.[1,2] Teaching about the optical properties of polymer films connects the above mentioned disciplines to chemistry and materials science and engineering, and to biology where molecular orientation plays role in many significant processes, like cell division.

Optics is widely taught in the undergraduate physics curriculum and is a growing part of our everyday life. Matrix methods are usually emphasized in specialized physics courses, and less often in the general physics course for science majors. For students not majoring in physics, this is often the only opportunity to get introduced to those methods and applied to Stokes analysis. Optical communications, LCD displays, modern telescopes and microscopes play major roles in how we obtain our information about the surrounding world. They strongly influence the development of science and technology. Integration of physics and mathematics curricula is beneficial, requiring a larger variety of mathematical methods to be used in physics. Stokes analysis unifies both the benefits of learning optics and computer programming, and exercising matrix calculations. Chemistry and biology labs already use polarization to determine the concentration of a solution of an optically active substance[3,4] without utilizing Stokes analysis.

## II. Stokes analysis in current research

The importance of Stokes analysis stems from its growing applications. It has been used to analyze polarization not only in lab settings but also for atmospheric aerosol,[5] near-field polarimetry[6] and for fiber-optic communication systems.[7-11] The interest in developing new instruments using Stokes analysis is increasing with new applications to spectroscopy.[12] Stokes analysis has been used in medicine to analyze adaptive optics images of the retina,[13] tomography.[14,15] It has been applied to semiconductor optical amplifiers,[16] backscattering experiments,[17] in astrophysics[18, 19] and for Cosmic Microwave Background measurements.[20] The method is an active area of theoretical investigation and improvement as well.[21] The Mueller Matrix for long period fiber gratings has also been derived.[22,23]

## III. THEORY OF THE METHOD

Here we show the classical method for defining the Stokes vector. The expressions for the elliptically polarized light with relative phase-shift δ are:[24]

$$E_x(t)=\mathbf{\hat{i}}E_{0x}(t)\cos[(kz-\omega t)+\delta_x] \qquad (1)$$

$$E_y(t)=\mathbf{\hat{j}}E_{0y}(t)\cos[(kz-\omega t)+\delta_y] \qquad (2)$$

where $E_x$ and $E_y$ are the mutually orthogonal components of the electric field, and $E_{0x}$ and $E_{0y}$ are their amplitudes, **î** and **ĵ** are unit vectors along x and y axes respectively, ω is the angular frequency, and $k=2\pi/\lambda$ is the propagation wavenumber, $\delta=(\delta_y-\delta_x)$ is the phase difference between the two components of the EM wave ($-180°<\delta<+180°$), where $\delta_y$ is the phase of the y projection of the electric vector, and $\delta_x$ is the phase of the x projection of the same vector. $\mathbf{k\cdot r}=$ kz for one dimensional propagation, where **r** is a radius vector and z is the propagation direction. Vector quantities are in bold while their components and magnitudes are not in bold.

Using the amplitudes of the components of the electric field vector, the Stokes polarization parameters can be written as:[24]

$$S_0=E_{0x}^2 + E_{0y}^2 \qquad (3)$$

$$S_1=E_{0x}^2 - E_{0y}^2 \qquad (4)$$

$$S_2=2E_{0x}E_{0y}\cos\delta \qquad (5)$$

$$S_3=2E_{0x}E_{0y}\sin\delta \qquad (6)$$

Stokes represented un-polarized, circular, and linearly polarized light by using optical elements to separate the components of the light.

The Stokes parameters form a vector called the Stokes vector:

$$S=\begin{pmatrix} S_0 \\ S_1 \\ S_2 \\ S_3 \end{pmatrix} \qquad (7)$$

Orientation is measured from the vertical X-axis. The Stokes parameters can be measured as:

$S_0$ - total intensity. $S_0=\langle I\rangle$.

$S_1$ - difference in the intensities, between orthogonal vertically and horizontally linearly polarized components at 0° and 90° ($I_0$-$I_{90}$). $S_1=\langle I_{0°}-I_{90°}\rangle$

$S_2$ - difference in the intensities, between orthogonal linearly polarized components at angles ±45° ($I_{-45}$-$I_{+45}$). $S_2=\langle I_{-45°}-I_{+45°}\rangle$.

$S_3$ - difference in the intensities, between right and left circularly polarized components ($I_{rcp}$-$I_{lcp}$). $S_3=\langle I_{rcp}-I_{lcp}\rangle$.

Where $\langle\rangle$ refer to statistical averaging of the intensities of the photons with particular polarization.

The normalized Stokes parameters (divided by the total intensity $S_0$) for light transmitted through different optical elements, form their Stokes vectors.[25,26] Some examples are:

$$S=\begin{pmatrix} 1 \\ 0 \\ 0 \\ 0 \end{pmatrix} - \text{Un} - \text{polarized light} \qquad (8)$$

$$S = \begin{pmatrix} 1 \\ 1 \\ 0 \\ 0 \end{pmatrix} \quad - \text{linearly polarized light at } 0°, ( I_0 = 1, I_{90} = 0 ) \tag{9}$$

$$S = \begin{pmatrix} 1 \\ -1 \\ 0 \\ 0 \end{pmatrix} - \text{linearly polarized light at } 90°, ( I_0 = 0, I_{90} = 1 ) \tag{10}$$

$$S = \begin{pmatrix} 1 \\ 0 \\ 0 \\ 1 \end{pmatrix} \quad - \text{Left circularly polarized light}, ( I_{rcp} = 0, I_{lcp} = 1 ) \tag{11}$$

$$S = \begin{pmatrix} 1 \\ 0 \\ 0 \\ -1 \end{pmatrix} \text{- Right circularly polarized state} ( I_{rcp} = 1, I_{lcp} = 0 ) \tag{12}$$

The Stokes polarization parameters are connected though the relationship $S_0^2 = S_1^2 + S_2^2 + S_3^2$. The degree of polarization Π can also be represented using Stokes analysis by $\Pi = [(S_1^2 + S_2^2 + S_3^2)/S_0^2]^{1/2}$. Π=1 for perfectly polarized light, Π=0 for perfectly un-polarized light, 0<Π<1 for partially polarized light.[27,28] This expression allows one to measure the degree of depolarization of light by a sample as D=1- Π.[29,30]

The power of Stokes analysis is that the operation of optical devices on light can be described by using their real 4x4 Mueller matrices. The Stokes nondiagonalizable Mueller matrix[31,32] can be used further to describe the effects of depolarization.

The Mueller matrices for the optical elements used in this lab are for a rotating linear polarizer and for a rotating retarder. The general form for a rotating linear polarizer, where γ is azimuthal angle of rotation of the polarizer is:

$$M = \frac{1}{2}\begin{bmatrix} 1 & \cos 2\gamma & \sin 2\gamma & 0 \\ \cos 2\gamma & \cos^2 2\gamma & \cos 2\gamma \sin 2\gamma & 0 \\ \sin 2\gamma & \cos 2\gamma \sin 2\gamma & \sin^2 2\gamma & 0 \\ 0 & 0 & 0 & 0 \end{bmatrix} \tag{13}$$

The transformation of the Stokes vector of the polarized light by the Mueller matrix of the optical element for Π=1is obtained by:

$$S_{final}=M.S_{initial} \tag{14}$$

The Mueller Matrix of any optical element can be measured experimentally.[33]

The action of an oriented polymer sample producing phase-shift δ, usually called optical retardation – R, where δ=R, on the polarized light is given by the matrix for a rotating retarder:[34]

$$\begin{vmatrix} 1 & 0 & 0 & 0 \\ 0 & \cos^2 2\gamma + \sin^2 2\gamma \cos R & \sin 2\gamma \cos 2\gamma (1-\cos R) & -\sin 2\gamma \sin R \\ 0 & \sin 2\gamma \cos 2\gamma (1-\cos R) & \sin^2 2\gamma + \cos^2 2\gamma \cos R & \cos 2\gamma \sin R \\ 0 & \sin 2\gamma \sin R & -\cos 2\gamma \sin R & \cos R \end{vmatrix} \tag{15}$$

The origin of the optical activity in our lab is explained by the polymer molecular polarizability which has different values in directions parallel and perpendicular to the axis of the molecule, as seen in Fig. 1. We use the phenomenon that in anisotropic polymer samples the components of the electromagnetic radiation propagate with different speeds along directions with different indices of refraction and have a new state of polarization when recombined after the sample (Fig. 2 and Fig. 6).

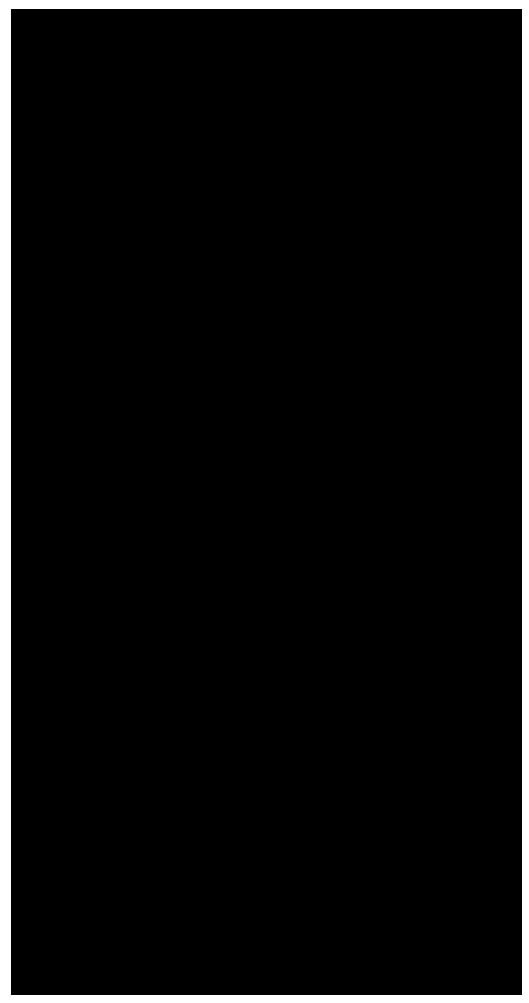

Fig. 1. This is an illustration of the origin of the optical anisotropy on molecular level for a polymer. The polymer chain is represented by the zig-zag line. The polarizability $\alpha_z$ along the z axis can be greater or smaller than those along the x or y axis depending on the polymer used [Redrawn from Viney 1990, Fig. 2.6] .[35]

The dependence of the refractive index on the molecular polarizability is:

$$(n^2-1)/(n^2+2)=N\alpha/3\varepsilon_0 \quad (16)$$

where n – refractive index, N – number of molecules per 1 $cm^3$ , and $\varepsilon_0$ - the free space permittivity. This relation is valid only for isotropic distribution of molecules with mean polarizability α.[24]

If there is anisotropy of the molecular polarizability Δα it leads to anisotropy of the index of refraction, which defines the birefringence:[24]

$$\Delta n=((n^2+2)/3)^2(N\Delta\alpha/2\varepsilon_0) \quad (17)$$

When the sample is not oriented, all of the indices of refraction are averaged, giving rise to a sample Fletcher indicatrix that is spherical in shape (Fig. 3). In this case the polarization state of the transmitted light is not changed after passing through the sample. When the sample is oriented, the molecules are statistically ordered preferentially in one direction, and the average index of refraction over all of the molecules in a small area of the sample has different values in parallel and perpendicular to the orientation directions. This deforms the spherical Fletcher indicatrix into an ellipsoid (Fig. 2).

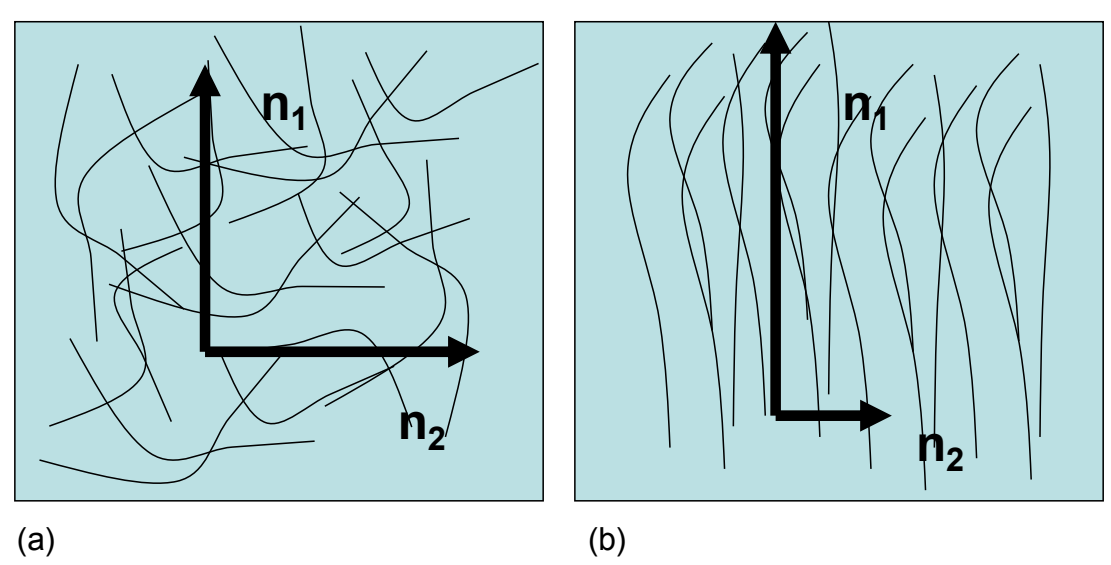

Fig. 2. An illustration of the molecular origin of the macroscopic optical properties of: (a) isotropic, unoriented polymer sample where $n_1=n_2$; (b) anisotropic, oriented polymer sample where $n_1 \neq n_2$. The curved lines represent the polymer chains. The arrows represent each of the indices of refraction, $n_1$ and $n_2$ as indicated.

On Fig. 2 the black curves represent the orientations of polymer chains in a thin, optically transparent polymer film. In the isotropic case (a) the molecular polarizabilities of the polymer chains are averaged in all directions, which statistically determines that the indices of refraction of the sample in two perpendicular directions are the same $n_1=n_2$. In the anisotropic case (b) the molecular polarizabilities of the polymer chains are not averaged in all directions, which statistically determines that the indices of refraction of the sample in two perpendicular directions are different $n_1 \neq n_2$ which defines $\Delta n = n_1 - n_2$ as the birefringence of the polymer sample, the difference in the index of refraction of the polymer in parallel and perpendicular directions to the anisotropy axis. The measured intensity after the analyzer then depends on the angle of orientation of the sample with respect to the direction of the initial linearly polarized light and the analyzer.

The intensity of the transmitted light is proportional to the phase-shift (optical retardation) R, $R = 2d(\Delta n)/\lambda$, where d is the thickness of the sample, $\lambda$ the wavelength of light and $\Delta n = n_1 - n_2$. It also depends on the initial intensity $I_0$ and azimuthal angle $\gamma$.

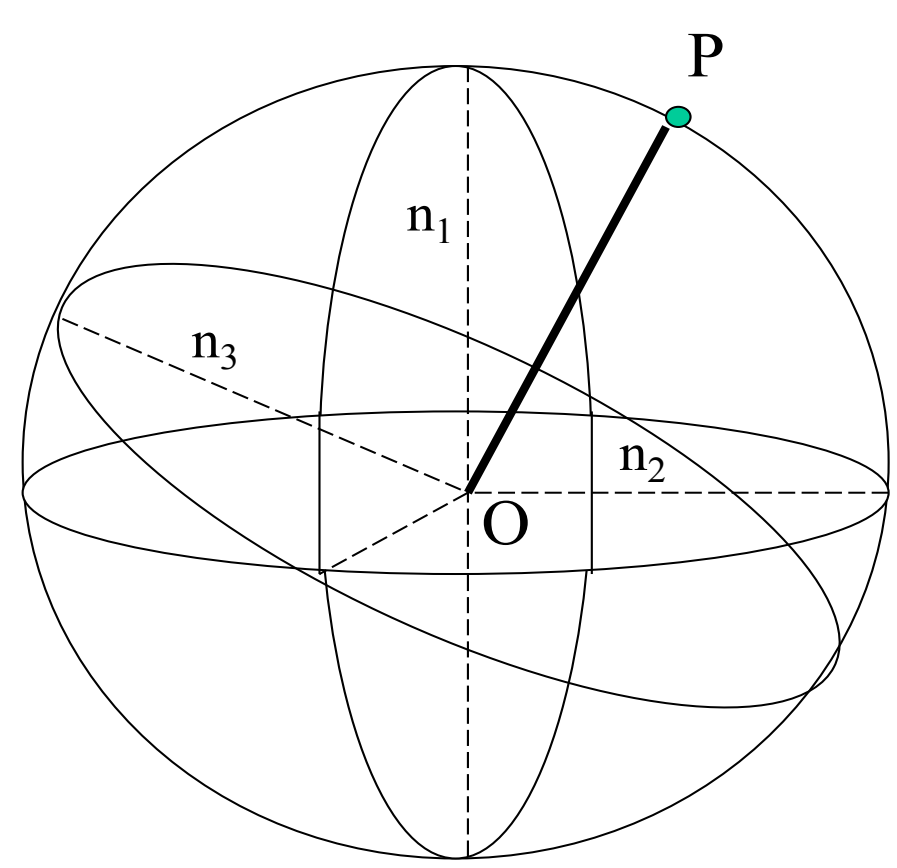

Fig. 3. This figure represents the isotropic Fletcher indicatrix formed by the indices of refraction of the unoriented isotropic sample. In this particular case the indicatrix is not an ellipsoid but a sphere, due to the identical index of refraction in all three directions. The vector of the electric field forms a circular cross-section with the sphere along any direction of propagation of light and the polarization of the light emerging after the sample is not changed. OP is the direction of propagation of the EM wave.

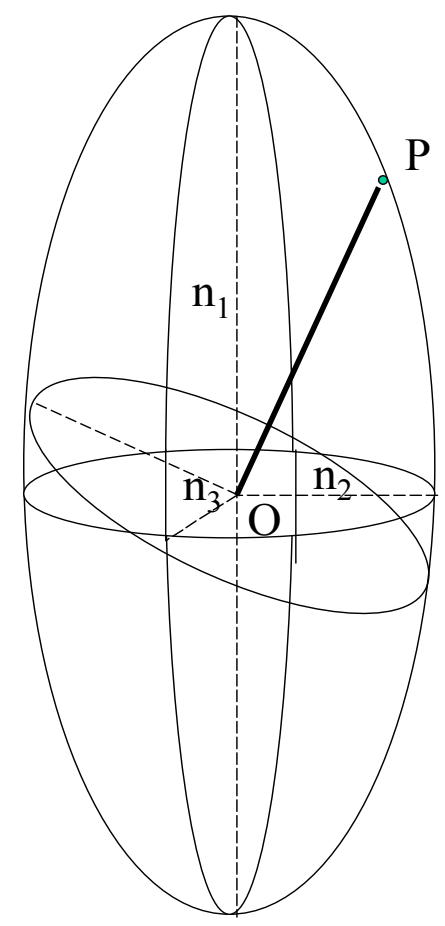

Fig. 4. The anisotropic Fletcher indicatrix formed by the indices of refraction of an oriented polymer sample is an ellipsoid formed by the different indices of refraction. OP is the direction of propagation of the EM wave. The distance to any point P on the Indicatrix is $n_{op}$, which is the radius-vector to that point. The light interacts with the ellipse formed by the cross-section of the electric field vector plane perpendicular to the propagation direction and the ellipsoid.

In our setup we orient the drawn uniaxial polymer sample with the long axis of the ellipsoid formed by the three indices of refraction in the sample perpendicular to the direction of propagation of light OP. In this orientation we are measuring the elliptical cross-section perpendicular to OP, shown on Fig. 5.

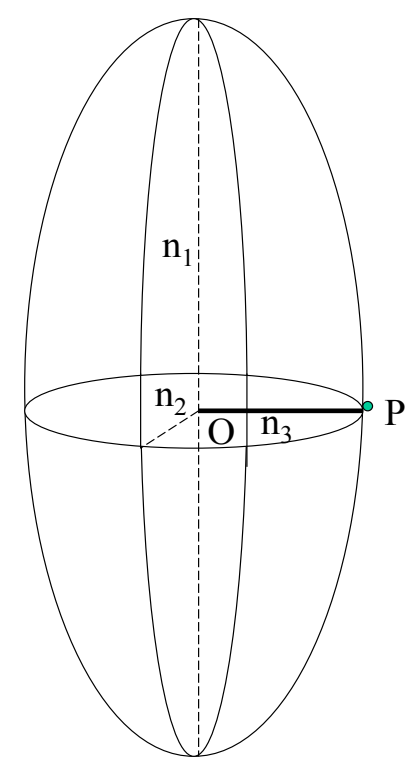

Fig. 5. On this figure we show the orientation of the Fletcher indicatrix formed by the anisotropy of the indices of refraction in our uniaxial sample, with respect to the direction of the propagation of light OP specifically for our experiment.

## IV. EXPERIMENTAL METHOD

Using matrix multiplication the students derive the form of the Stokes vector for the particular state of the polarized light in their setup based on the orientation of the polarized laser and the chosen coordinate system. They write computer programs to help them with the calculation throughout the lab. Using the general matrix of a linear polarizer they derive its form at different rotation angles of the analyzer used during the lab. Using the matrix of a general retarder, the students derive its form for the particular optical activity and rotation of the polymer sample. Then by matrix multiplication they obtain an expression for the expected intensity at the detector. They compare their theoretical calculations with their experimental measurements in order to verify the method of the Stokes analysis and draw conclusions. After that, the students compare the intensity measured upon rotation of the analyzer: 1. in a setup without a sample; 2. with an isotropic (non-birefringent) sample; 3. with an anisotropic birefringent sample. In the first setup without the sample the students derive the familiar to them Malus's law theoretically using the Stokes analysis and experimentally verify it to convince themselves in the power of the Stokes method.

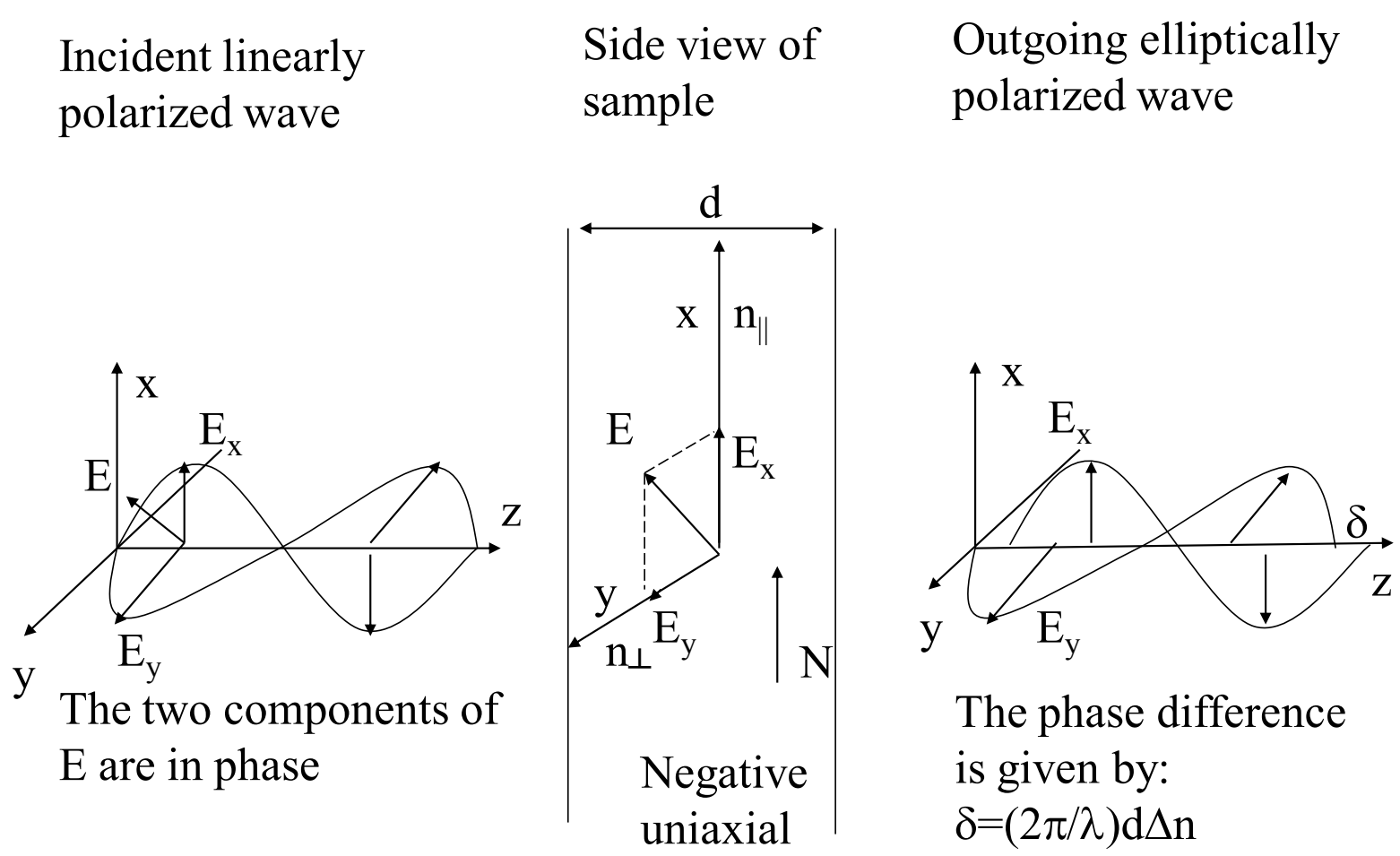

Fig. 6. This figure represents the phase-shift of linearly polarized light leading to elliptically polarized light introduced by an optically active sample. λ is the wavelength, d is the thickness of the polymer film, Δn is the birefringence.

If the lab is extended to two weeks, the students can further their understanding of anisotropic optically active materials by connecting the measurements of the intensities at several different angles with the phase-shift of the light passing through the sample to draw quantitative conclusions about the birefringence of the sample. For that they will need to measure its thickness and to know the wavelength of the laser. They can connect the measured birefringence to the internal order parameter in the sample and the polarizability of the molecule (Fig. 1).

During this lab the students gain practical experience with the laser alignment of the system. They also understand the mathematical representation of polarized light, polarizing optical elements, the sample and the intensity of the polarized light, constructing mathematically a polarizing optical system and finding out about the internal order of an optically active polymer sample.

**Part 1: Deriving Malus's law**

This section serves to demonstrate to the students the usefulness of the Stokes method in analyzing the polarization of the light through an optical system, by using it to derive the well known Malus's law.

$I(\gamma)=I(0)\cos^2 \gamma$

Where $I(\gamma)$ is the intensity of the light at an angle between the polarizers, $\gamma$. I(0) is the initial intensity of the light, identical to the case when $\gamma=0$.

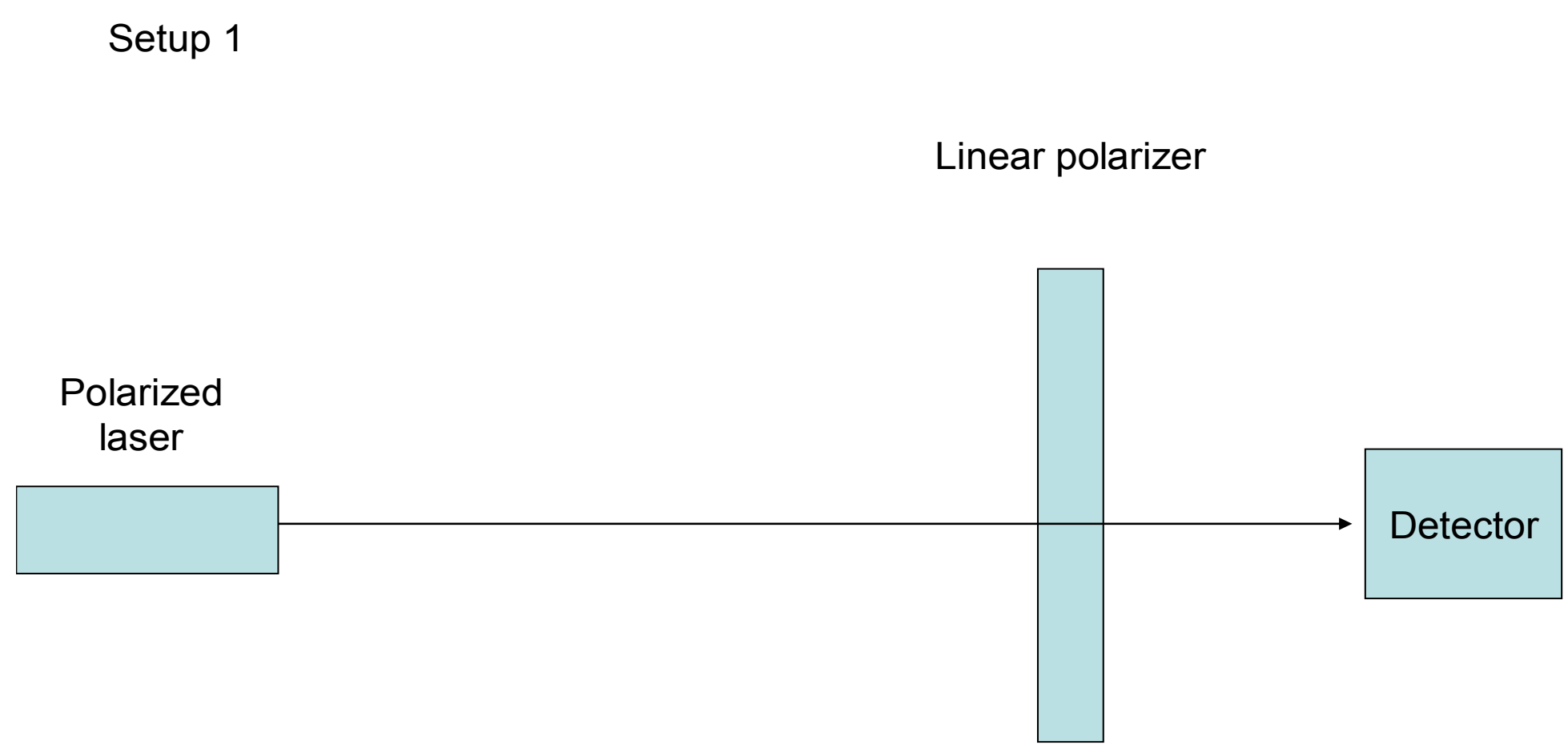

Fig. 7. Experimental setup for the first part of the lab. It consists of three elements on an optical rail: vertically oriented linearly polarized laser; initially vertically oriented transmission axis of a linear polarizer and a light intensity detector.

The students learn how to apply the method, gain conceptual and mathematical understanding of it, and are convinced of its utility. This prepares them to apply it to later parts of the lab for unknown samples.

The setup consists of initially vertical linearly polarized laser light with $\lambda$=633nm being sent through a linear analyzer and then the transmitted intensity being captured by a light detector. In order to learn laser alignment on an optical bench the students use laser reflections to make sure that all of the elements are oriented at 90° with respect to the optical path.

The students first perform Stokes analysis and fill in a table for calculated predictions for the intensity of the light as a function of the angle of the orientation of the linear polarizer. They use the measured initial intensity $I_0$ when the orientation of the incident polarized light and the transmission axis of the linear polarizer are parallel. Then the students turn the analyzer at 90° and measure the intensity, explain why it is not zero, and use it as a constant background intensity $I_b$ to subtract from all of the intensities measured throughout the lab. Using their measurement of $I_0$ they make a prediction by calculating the intensity of the light at the detector when the orientation of the linear polarizer is changed in increments of 10° from 0° to 90°. As an exception the students calculate the predicted intensity at 45°. Once they have filled in a table, the students graph their results. After they are done with the calculation part the students rotate the analyzer and fill in the values for the intensity measured at each orientation with the background intensity subtracted and graph their experimental result. In the last column of the table they find the percent difference between the predicted and measured intensities. To avoid large errors they do a linear fit of $\cos^2 \gamma$. After the students have finished their measurements they compare the two graphs and discuss their observations.

**Part 2: Double refraction: Un-oriented polymer sample**

Before the students insert the sample they record the reading for $I_0$ when the analyzer (same from part 1) is oriented at 0º - maximum intensity and then carefully center the un-oriented polymer sample. They record the power meter reading as $I_0$ and rotate the polymer sample to 45º and to 90º, recording the intensities $I_{45}$ and $I_{90}$ discussing their results and the reduction of intensity due to sample scattering. The students explain the observed effect through its molecular origins as in Fig. 2. The students repeat the steps just described with the analyzer at 90º. In this case without the sample the intensity is at minimum - $I_b$. They print and explain the intensity vs angle graph.

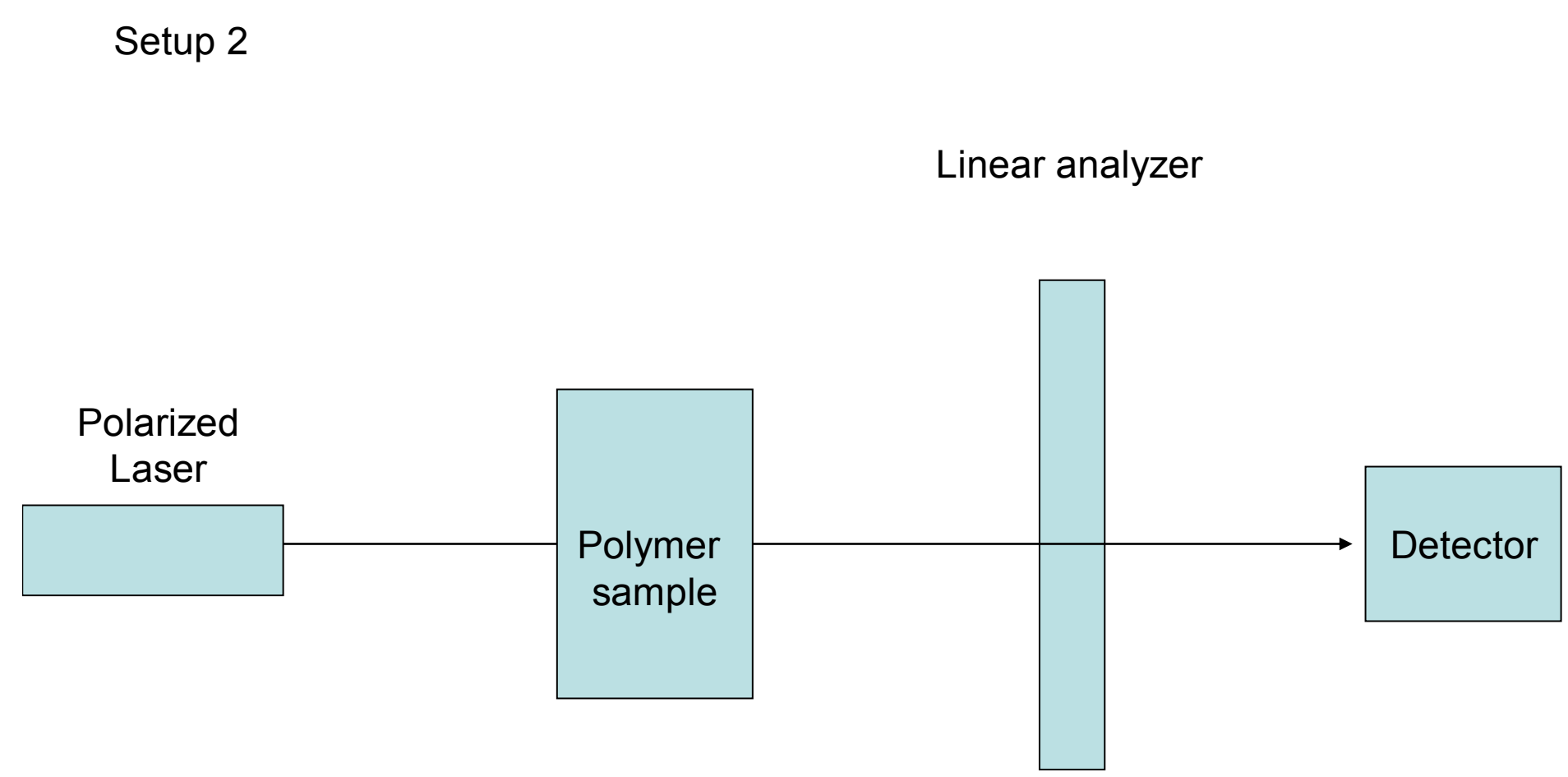

Fig. 8. An experimental setup for parts 2 and 3 of the lab. It consists of four elements: vertically oriented linearly polarized laser; a polymer sample; a linear analyzer set at 0° and 90° in respect to the initial linearly polarized light from the laser; a light intensity detector.

**Part 3: Double refraction: Oriented polymer sample**

Next the students replace the isotropic with anisotropic polymer sample with unchanged setup from last experiment where the analyzer is oriented at 90º and repeat the measurements. When the oriented polymer sample is at 0º the power meter reading are recorded. Then the students rotate the polymer sample to 45º and to 90º, and record the $I_{45}$ and $I_{90}$ and discuss the results using Fig. 1 and Fig. 2.

On the graph intensity vs angle the students make an observation about what is the shape of the curve and at what orientations the intensity has a maximum and a minimum and why. They reach the conclusion that the curve is sinusoidal with maxima at n*45º, where n=1,3,5,7, and minima at n*90º, where n=0,1,2,3.

The intensity is unchanged at $I_b$ for sample orientations at 0º and 90º. At those two particular angles the linearly polarized light is oriented exactly along $n_1$ and $n_2$, and it travels with the same speed through the sample. There is no projection of the vector of the linearly polarized light along the other direction and the polarization state of the light is not affected. For any other angle there are projections of the linearly polarized vector along the two indices of refraction, and the intensity of the transmitted light increases, with a maximum when the sample is at 45º. It is straightforward for the students to measure with micrometer the thickness of the sample and use the expression for phaseshift to estimate the birefringence of the sample.

**CONCLUSIONS:**

Accounting for the specifics of the general undergraduate Physics curriculum we have constructed a lab which ties the material taught in lecture with the hands-on activities in the lab, and also introduces Stokes analysis as an efficient method to calculate the intensities of the polarized light transmitted through anisotropic polymer sample. This lab uses our previous research in developing a microscopic transmission ellipsometric system for research in polymer physics and adapts it to the level of knowledge of students enrolled in general physics. Since most of the students in this course are not physics majors, learning about optical activity in polymers connects physics to other disciplines, like chemistry, biology and engineering, which is increasingly important. This lab also ties physics to mathematics and computer science by asking students to apply and develop their knowledge of matrix calculations and computer programming which is usually done in advanced physics and optics classes for physics majors but is also important for all science majors taking general physics. This knowledge prepares the students for real world optical applications after they graduate and enter the workforce. The students responded positively to the new lab and have completed it in the allocated time period.

**ACKNOWLEDGEMENTS**

The authors thank Assumption College for the professional environment making this work possible; Martha Slavkovsky and Erin Gombos for helping with revisions of the draft; the students in General Physics II and Honors General Physics II courses who worked through preliminary versions of this laboratory.

[a]Electronic mail: ggeorgie@assumption.edu; georgi@alumni.tufts.edu